\documentclass{WileyMSP-template}

\begin{document}


\title{Flexible hydrogels connecting adhesion and wetting}

\maketitle


\author{A-Reum Kim$^{a}$},
\author{Surjyasish Mitra$^{a}$},
\author{Sudip Shyam},
\author{Boxin Zhao*}, and
\author{Sushanta K. Mitra*}



\begin{affiliations}
Dr. A-Reum Kim, Prof. Boxin Zhao*\\
Department of Chemical Engineering, Waterloo Institute for Nanotechnology, University of Waterloo, Waterloo, Ontario N2L 3G1, Canada.\\
Email Address: zhaob@uwaterloo.ca

Dr. Surjyasish Mitra, Dr. Sudip Shyam, and Prof. Sushanta K. Mitra*\\
Department of Mechanical \& Mechatronics Engineering, Waterloo Institute for Nanotechnology,
University of Waterloo, Waterloo, Ontario N2L 3G1, Canada.\\
Email Address: skmitra@uwaterloo.ca

\end{affiliations}

$^{a}$ These authors contributed equally. 


\keywords{Adhesion, Wetting, Hydrogels, Soft matter}

\begin{abstract}
\justifying
Raindrops falling on window panes spread upon contact, whereas hail can cause dents or scratches on the same glass window upon contact. While the former phenomenon resembles classical wetting, the latter is dictated by contact and adhesion theories. The classical Young-Dupré law applies to the wetting of pure liquids on rigid solids, whereas conventional contact mechanics theories account for rigid-on-soft or soft-on-rigid contacts with small deformations in the elastic limit. However, the crossover between adhesion and wetting is yet to be fully resolved. The key lies in the study of soft-on-soft interactions with material properties intermediate between liquids and solids. In this work, we translate from adhesion to wetting by experimentally probing the static signature of hydrogels in contact with soft PDMS of varying elasticity of both the components. Consequently, we probe this transition across six orders of magnitude in terms of the characteristic elasto-adhesive parameter of the system. In doing so, we reveal previously unknown phenomenology and a theoretical model which smoothly bridges adhesion of glass spheres with total wetting of pure liquids on any given substrate. Lastly, we highlight how solid like hydrogels can be treated as potential candidates for cleaning impurities from conventionally sticky PDMS substrates.

\end{abstract}


\section{Introduction}
\justifying
Soft polymeric materials like hydrogels and elastomers find a wide range of applications, both in advanced technologies and everyday life. Hydrogels are extensively utilized in biomedical engineering, including in-vitro model studies \cite{matsuzaki2014high}, 3D bio-printing inks \cite{zhou20233d}, and drug delivery \cite{alici2015towards}, due to their biocompatibility, biodegradability, and high-water content \cite{yang2020hydrogel}. 
Compared to hydrogels, elastomers possess greater stability in diverse environments and mechanical robustness, unless specially engineered tough hydrogels are considered \cite{stricher2015met,yuk2016skin}. Consequently, elastomers are employed in applications that require enhanced stretchability and durability, such as soft robotics \cite{shintake2018soft,zhang2021inchworm}, microfluidic devices \cite{fujii2002pdms,lamberti2014pdms}, and sensors \cite{kim2019soft,guo2016highly}. 
Hydrogels and elastomers possess distinct characteristics and often utilized to compensate for each other's weaknesses \cite{yuk2016skin,demirci2022elastomer,grosjean2022bioresorbable,le2017wearable} . The interaction between these soft materials does not adhere to either traditional wetting or solid contact mechanics, necessitating a comprehensive understanding of the interplay between their wettability and rheological properties to optimize their performance across different applications.

Contact between two elastic surfaces dates
back to the seminal work by Hertz elucidating
the morphology of the contact zone \cite{hertz1882beruhrung}.
Briefly, for two surfaces with radii, elastic
moduli, and Poisson's ratios $R_{1}$, $E_1$, $\nu_{1}$ 
and $R_{2}$, $E_2$, $\nu_{2}$,
the contact radius scales as, 
$a \sim \left(3RF/4E^{*}\right)^{1/3}$,
where $F$ is the applied load. 
Here, $R =\left(1/R_{1} + 1/R_{2}\right)^{-1}$
and $E^{*} =\left[(1-\nu_{1}^2)/E_{1} + (1-\nu_{2}^2)/E_{2}\right]^{-1}$
are the effective radius and effective elastic modulus, respectively \cite{hertz1882beruhrung}.
Though Hertz theory successfully explains
non-adhesive contacts, it fails where adhesion is present.
Taking \emph{adhesion} between the contacting surfaces
into account, Johnson-Kendall-Roberts (JKR) \cite{johnson1971surface}
and Derjaguin–Muller–Toporov (DMT) \cite{derjaguin1975effect} 
proposed modified theories. 
Considering adhesion, using JKR model, the modified contact radius 
can be expressed as \\
$a \sim \left[\left(3R/4E^{*}\right)\left(F+3\pi wR+\sqrt{6\pi wRF +(3\pi wR)^{2}}\right)\right]^{1/3} $, 
where $w = \gamma_{1} + \gamma_{2} - \gamma_{12}$
is the work of adhesion \cite{johnson1971surface}. 
Here, $\gamma_{1}$ and $\gamma_{2}$ are the surface tensions of the individual contacting pairs, whereas $\gamma_{12}$ is the interfacial tension.
Both JKR and DMT theories have been successfully adopted
to explain contacting surfaces \cite{butt2005force,style2013surface,liu2006deformation,
prokopovich2011adhesion,erath2010characterization}. 
However, they possess limitations too \cite{butt2005force,style2013surface,maugis1995extension,
pham2017elasticity}. 
For instance, for contacting surfaces with large contact radius, 
i.e., $a/R >> 0.1$,
the Maugis model has been found to provide more accurate 
description of the contact zone \cite{maugis1995extension,maugis1992adhesion}. 
Further, recent experiments involving glass spheres 
on very soft gels ($E^{*} \sim 1\,{\rm kPa}$) have demonstrated
how the JKR theory breaks down
due to the dominant role of solid surface tension \cite{style2013surface,pham2017elasticity}. 
Also, soft gels in contact with a glass sphere can often undergo long-term deformation dynamics  
which disrupts our conventional understanding of contact mechanics \cite{kim2022capillary}.
It appears that an universal model is lacking to explain contact between
any two surfaces. 

On the other extreme, wetting of liquids \cite{young1805iii,de1985wetting,de2004capillarity} essentially corresponds to $E^{*} = 0$. Existing literature has attempted to bridge contact mechanics with a finite $E^*$ to wetting with $E^{*} = 0$ using theoretical treatment \cite{salez2013adhesion,long2016effects,zhang2019refined,xu2014effects,
hui2015indentation,cao2014adhesion}. 
However, most of them are either restricted to small-strain
systems or consider the variation of elasticity of only one
of the contacting pairs. 
Similarly, existing experimental studies are limited in scope. 
Most experiments consider rigid-on-soft \cite{butt2005force,style2013surface,pham2017elasticity,
kim2022capillary,xu2014effects,hui2015indentation,cao2014adhesion,rimai1989adhesion,rimai1994adhesion,ally2010interaction,
jensen2017strain,jensen2015wetting} or soft-on-rigid \cite{liu2006deformation,salez2013adhesion,liu2021capillary,
sokoloff2016effects,chakrabarti2018elastowetting} contacts.
Hence, the effective elastic modulus $E^{*} = E_{1}/(1-{\nu_{1}}^{2})\, (\rm{or}\,E_{2}/(1-{\nu_{2}}^{2}))$, i.e., always
becomes a function of only one of the elastic modulus, and thus fails to provide a holistic picture. 
Soft-on-soft systems provide the ideal playground to
tune the effective elastic modulus $E^{*}$, taking into 
account, elasticity of the contact pair and provide
an excellent platform to investigate the transition.
At the same time, many biological \cite{schwarz2013physics} and artificial materials \cite{zhao2017electronic} represent 
soft-on-soft contact systems. 
Understanding such systems is key to interpreting natural phenomena like cell adhesion \cite{schwarz2013physics,gong2018matching,li2008afm} as well as in developing advanced technologies like medical diagnostic chips \cite{regehr2009biological}, contact lenses \cite{kopecek2009hydrogels}, and artificial organs \cite{zhao2017electronic}.

\section{Results and Discussions}
\subsection{Flexible hydrogels}
As shown in Fig.~\ref{fig:1}a, we prepared hydrogel drops using acrylamide (AAm) as a monomer, N,N'-Methylene-bis-acrylamide (BIS) as the crosslinker, and 2,4,6-tri-methyl benzoyl-diphenylphosphine oxide (TPO) nanoparticle as the initiator. 
To prepare the spherical shape of hydrogel, we used the density gradient between octane and silicon oil.
We pipetted 4\,$\mu$L of hydrogel in the middle of the two solvents and subsequently the pregel solution was UV cured. 
Post polymerization, the hydrogels were washed in heptane.
The elasticity of the hydrogels (i.e., $E_1$) is varied between 0.0057\,kPa to 392.80\,kPa 
by diluting the monomer in weight percentages of 4.0 - 30.0. 
We further prepared a \emph{liquid hydrogel} using 2.5 wt. \% of the monomer which behaves almost like a liquid with viscosity, $\mu_{1} = 0.163\,{\rm{Pa.s}}$ (Fig.~\ref{fig:1}b). It should be noted here that PAAm 4.0\% also behaves like a liquid  but for the sake of simplicity we will refer to PAAm 2.5\% as the liquid hydrogel. 
On the other hand, the stiffer hydrogels, i.e., PAAm 13\, wt\,\% onward resembles solid like contact morphology.
Thus, by varying the monomer wt. \%, we are able to mimic fully wetting liquid drops as well as non-wetting solid spheres exhibiting a broad control of wettability and elasticity (Fig.~\ref{fig:1}b). The contact/wetting morphology of hydrogels are similar for rigid glass substrates as well as soft substrates with a finite elasticity $E_{2}$. 
To prepare soft substrates, we used a combination of two different polydimethylsiloxane (PDMS) : Sylgard 184 and Sylgard 527 in weight ratios of 10:1 and 1:1, respectively. 
Consequently, we mixed the two solutions in different weight ratios to obtain soft substrates 
with Young's modulus (i.e., $E_{2}$) varying between 3\,kPa to 6855\,kPa. 

The non-wetting characteristics of stiffer hydrogels on soft PDMS
present an interesting consequence: they can freely roll off such soft substrates and in that process provide efficient surface cleaning (Fig.~\ref{fig:1}c). 
Even though water serves as the model liquid for self-cleaning, superhydrophobic surfaces, they are inappropriate candidates for soft substrates: the wetting ridge formation at the three phase contact line renders them sticky \cite{style2013universal,marchand2012contact,
jerison2011deformation}.
Despite their high water content, stiff hydrogels exhibit no such features and are potentially perfect candidates for cleaning PDMS based substrates used widely in digital microfluidics and biomedical devices. 
Additionally, spectroscopy measurements indicate that the prepared hydrogels exhibit negligible dehydration as well as diffusion of any constituent species upon contact with PDMS (Fig.~\ref{fig:1}d). Even plasma treatment of
PDMS bears no change in the non-wetting characteristics of the stiffer hydrogels on soft PDMS (Fig.~\ref{fig:1}e).
We note that enhancing surface wettability by plasma treatment makes water drops fully wet PDMS and thus makes them even more undesirable candidates for surface cleaning. 

At the same time, the extensive elasticity variation of both the top (hydrogel) and bottom (PDMS) contacting pair provides us  with the unique opportunity to 
probe adhesion-to-wetting transition. 
Consequently, we performed a thorough static characterization of 1\, mm radius hydrogels ($R_{0} = 0.96\pm 0.04\,\rm{mm}$) on 1\,mm thick, soft PDMS substrates across this wide range of both hydrogel ($E_{1}$) and PDMS ($E_{2}$) elasticity using shadowgraphy. 
For comparison, we also performed experiments of water droplets and glass spheres on the soft PDMS and rigid glass substrates. 
Further, we tested both pristine and plasma treated PDMS substrates to probe the influence of substrate wettability.
Assuming $\nu_{1}, \nu_{2} \approx 0.5$ \cite{kim2022capillary,chakrabarti2018elastowetting}, 
we thus vary $E^{*}$ between 0.007\,kPa to $10^{4}$\,kPa: a parameter space significantly broader than those probed in existing literature.

\subsection{Morphology of the contacting pairs}

In Fig.~\ref{fig:2}a, we recreate the different experimentally observed static configuration of spherical hydrogels on pristine, soft PDMS substrates and glass substrates using appropriate schematics. 
From Figs.~\ref{fig:1}b and \ref{fig:2}a, it is evident that 
the hydrogels form a distinct foot-like region close to the three phase contact line: a phenomenon first predicted
by Joanny et al. \cite{joanny2001gels} and later experimentally verified \cite{chakrabarti2018elastowetting}.
The foot region exhibits a finite contact angle $\theta^{*}$ 
which is markedly different than that exhibited by pure water/glycerol droplets on rigid or soft substrates (Figs.~\ref{fig:2}b,c). 
The appearance of the foot diminishes with decreasing hydrogel elasticity and disappears for the two liquid-like hydrogels (PAAm 2.5\,wt.\% and PAAm 4.0\,wt.\%)  mimicking wetting of pure liquids with a macroscopic apparent contact angle $\theta_{\rm m}$ (Figs.~\ref{fig:2}a-c). 
The foot region also diminishes for the stiffer hydrogels and although they possess a finite $\theta^{*}$, it is difficult to extract their foot angle with high accuracy due to very small foot dimensions. Consequently, for stiffer hydrogels, we use $\theta_{\rm m}$ to analyze their contact behavior. Thus, we show how with increasing elasticity $E_{1}$, the hydrogel contact angles increases from $\theta \rightarrow 0$ exhibited by complete wetting systems and eventually converges to $\theta \rightarrow 180^{\circ}$, commonly exhibited by glass spheres on any given surface (Fig.~\ref{fig:2}b). 
Further, we observe that for relatively stiffer hydrogels,
the variation of contact angles with substrate elasticity is insignificant whereas for the relatively softer ones, the observed
contact angles are higher on PDMS than glass substrates. 
As mentioned previously, plasma treatment of PDMS bears no effect
on the observed contact angles for stiffer hydrogels (Fig.~\ref{fig:2}d). 

As shown in Fig.~\ref{fig:2}e, irrespective of substrate elasiticity,
the hydrogels exhibit wetting to non-wetting characteristics with increasing elasticity $E_{1}$: a behavior manifested by 
gradually decreasing contact radius $a$. 
On one hand, the hydrogels with the two lowest elasticity behave like liquid drops and exhibit significant spreading upon initial contact with the substrates.
On the other hand, with increasing elasticity, the hydrogels exhibit decreasing spreading trend
and converges to the contact behavior shown by rigid glass spheres. 
However, the contact behavior, i.e., the contact radius is dominated mostly by hydrogel elasticity and shows certain subtle changes with substrate elasticity, which will be discussed later in text.  
To further analyze the contact morphology, we extract the apparent contact radius $a_{0}$ and apparent indentation depth 
$\delta$ by fitting the largest possible circle to the hydrogel profile (Fig.~\ref{fig:2}a).
Consequently, we highlight the deformation zone, i.e., the 
foot in terms of its height $h$ and length $l$ (Fig.~\ref{fig:2}a).
It can be observed from Fig.~\ref{fig:2}f that with increasing hydrogel elasticity, the foot-height reduces as well.
The foot height also diminishes in the low elasticity limit and completely disappears for the \emph{liquid hydrogel}. 
It should be noted here that the occurrence of foot is
primarily due to accumulation of polymers of the hydrogel \cite{joanny2001gels,bouillant2022rapid}.
A similar trend, i.e., decreasing with increasing elasticity is observed for the variation of apparent indentation depth and foot-length. 
Lastly, effect of plasma treatment bears negligible change to this individual parameter trends.

\subsection{Variation of contact radius with effective elastic modulus}
In Fig.~\ref{fig:3}a, we analyze the variation of the hydrogel-PDMS substrate as well as hydrogel-glass substrate contact radius $a$ for different hydrogels with the effective elastic modulus of the contacting pair $E^{*}$. 
It should be noted here that the contribution of surface (PDMS/glass) elasticity, i.e., $E_{2}$ is embedded in $E^{*}$. 
We observe that with increasing $E^{*}$, the contact radius
decreases and eventually converges to magnitude for rigid glass spheres atop the soft substrates. 
Interestingly, we observe that the hydrogel contact radius 
increases only marginally on plasma treated soft substrates (Fig.~\ref{fig:3}a, open symbols), exhibiting the same overall trend
with $E^{\ast}$ as before. 
The effect of underlying substrate
only becomes prominent for the softest hydrogel ($E_{1} = 0.0057\,{\rm {kPa}}$), where the contact radius is significantly larger for glass slides and plasma treated PDMS substrates as compared to pristine PDMS substrates.
This observation reflects that below a certain elasticity,
the hydrogels start to exhibit liquid-like wetting behavior:
higher degree of wetting for hydrophilic (glass, plasma treated PDMS) surfaces than hydrophobic (pristine PDMS) ones. 
At the same time, above a certain elasticity, i.e., for the two stiffest hydrogels, we observe higher contact
radii on glass and plasma treated PDMS than bare PDMS.
Thus, we observe deviations from wettability independence 
at both high and low elasticity limits.  

It should be noted here that for
relatively softer hydrogels, $a \approx R_{0}$ while 
for relatively stiffer ones, $a < R_{0}$. 
Consequently, the effective strain of the system,
i.e., $a/R_{0}$ varies from 0.05 to greater than 5.
Nevertheless, we first compare our experimental
data with Hertz and JKR theories. Note that for fitting
Hertz (or JKR), we have used $F = mg$, i.e., weight of the hydrogel.
We observe that Hertz theory deviates significantly
from our experimental data, whereas
JKR theory finds agreement only for certain cases.
Notably, for hydrogels and glass spheres, JKR theory provides
a satisfactory match for $E^{\ast} > 20\,\rm{kPa}$. 
This condition coincides with $a/R_{0} < 0.4$,
an important consideration for JKR (or Hertz) theory to hold.
Here, for fitting JKR theory, we used $w \approx 110 - 130\,\rm{mN/m}$ and $w \approx 44\,\rm{mN/m}$
for hydrogel-PDMS/glass substrates and glass-PDMS substrate \cite{kim2022capillary} contacts, respectively. 
It should be noted here that the lack of agreement
for Hertz theory is likely due
to presence of adhesion and large deformations present in our system. Also, since the Tabor parameter \cite{tabor1977surface}  relevant to the problem, $\beta = \left(R_{0}w^2/E^{*2}z_{0}\right)^{1/3} >> 1$, no comparison with DMT theory \cite{derjaguin1975effect} was performed. Here, $z_{0}$ is the characteristic length scale representing the range of surface forces and is in the order of nanometers \cite{liu2006deformation,tabor1977surface}. 

The limiting case of water droplets and \emph{liquid hydrogel} on the soft substrates
are shown in Fig.~\ref{fig:3}b which corresponds to $E^{\ast} \approx 0$. Consequently, we attempt to fit our hydrogel contact radius data occurring beyond the Hertz/JKR limit with that of wetting of pure liquids (water and {\emph{liquid hydrogel}). For the sake of completeness, we used different fitting laws: $a \sim E^{*-1/3}$ (JKR), $a \sim E^{*-1/4}$ (simplified Maugis \cite{chakrabarti2018elastowetting}), and $a \sim E^{*-1/5}$.  
Note that whereas the prefactor $C_{0} \sim w^{1/3}{R_{0}}^{2/3}$ can be derived from JKR theory, the prefactors $C_{1} \sim w^{1/4}{R_{0}}^{3/4}$ and $C_{2} \sim w^{1/5}{R_{0}}^{4/5}$ are obtained using dimensional analysis. 
For example, $w^{1/5}{R_{0}}^{4/5} \approx 0.0026\,{\rm N^{1/5}}{\rm m^{3/5}} \approx 0.7\,{\rm N^{1/5}}{\rm {mm}^{3/5}}$ (adjusted for units in $E^{*}$ converted to N/${\rm {mm}^{2}}$).
We observe that both the power laws with exponents -1/4 and -1/5 agree well with our experimental data whereas the JKR exponent -1/3 deviates significantly. 
Here, we note that both the scaling laws predict $a \to \infty$ as $E^{*} \to 0$. However, pure liquids exhibit finite wetting radius for partially-wetting and hydrophobic surfaces and only for complete wetting they form an extended thin film whose wetting radius can be considered approaching a very large value. Thus, the scaling laws are restricted only for \emph{total wetting} scenario. 
At the same time, in order to interpret the adhesion-to-wetting transition, we need to analyze our experimentally observed contact radius with respect to the
relative significance of work of adhesion $w$ which promotes contact and effective elasticity $E^{*}$ which opposes it.
Thus, to better represent our entire range of experimental parameters,
we non-dimensionalize our observed contact radius $a$ with respect to hydrogel radius $R_{0}$ and the relative significance of adhesion energy and elastic energy in the form of non-dimensionalized elasto-adhesive parameter $E^{\ast}R_{0}/w$ \cite{zhang2019refined,hui2015indentation}.  
Here $w$ accounts for all the relevant work of adhesion present in the system i.e., glass-PDMS, hydrogel-PDMS, and hydrogel-glass. 
Additionally, using non-dimensional analysis aids us in effectively comparing our findings with similar phenomena studied in existing literature \cite{style2013surface,rimai1989adhesion,rimai1994adhesion}.

\subsection{Transition between wetting and adhesion}
Here, we analyze the variation of normalized contact radius 
$a/R_{0}$ with non-dimensionalized elasto-adhesive parameter $E^{\ast}R_{0}/w$.
We compare our experimental data with the
normalized form of JKR as well the Maugis theory.
Maugis' theory \cite{maugis1995extension,maugis1992adhesion,maugis2000contact} is applicable for large deformations and a wide parameter space.
According to this approach, the contact radius can be
expressed as \cite{maugis1995extension}, 
\begin{equation}
F = \frac{3aE^{*}}{2}\left(\frac{{R_{0}}^2+a^2}{4a}\ln{\frac{R_{0}+a}{R_{0}-a}} -\frac{R_{0}}{2}-\sqrt{\frac{8\pi aw}{3E^{*}}}\right),
\label{eqn:1}
\end{equation}
where $F$ is the applied load. 
On numerically solving Eq.~\ref{eqn:1} using our system properties, 
we observe that \emph{real number} solutions are obtained only intermittently. Note that while fitting JKR and Maugis, using $F = 0$ or $F = mg$ bears negligible difference on the final solution since the radius
of the hydrogel is below the capillary length-scale.
Further, since our experimental contact radius
spans across a large range, i.e., 0.06\,mm - 5\,mm and assumes values close to and even greater than the hydrogel radius of 0.96\,mm, we approximated Eq.~\ref{eqn:1}
for the case $a \to R_{0}$, yielding $a \sim (\pi w/E^{*})$, or equivalently $a/R_{0} \sim (\pi E^{*}R_{0}/w)^{-1}$.
As shown in Fig.~\ref{fig:4}a, normalized JKR theory overlaps with some of our experimental data for $E^{*}R_{0}/w > 200$ whereas both general (Eq.~\ref{eqn:1}) and large strain approximation of the Maugis' theory intermittently obey our experimental data. 
Consequently, we observe that most of our experimental
data can be fitted using the scaling law, $a/R_{0} \sim \left(E^{\ast}R_{0}/w\right)^{\alpha}$, where the exponent $\alpha$ can be either of -1/4 or -1/5.
Here, the fitting parameter is of the order of unity.
However, upon plotting existing literature data corresponding to rigid-on-soft as well as soft-on-rigid contact systems in the same plot, we observe that the scaling law with exponent $\alpha = -1/4$ is in better agreement. 
In this regard, we note that even though majority of literature data represents contact of rigid spheres on soft substrates, they fall nicely on our experimental data, validating our approach to use the elasto-adhesive parameter to represent such systems.

Although, both the aforementioned scaling laws reasonably agree with our experimental data across six order of magnitude in $E^{*}R_{0}/w$, we argue that $E^{*}R_{0}/w \approx 200$ can be thought of as a reasonable transition point above which JKR theory is applicable considering experimental bounds. 
Incidentally, using this limit in the JKR theory predicts a strain of $a/R_{0} = \left(9\pi/400\right)^{1/3}\approx 0.4$, which coincides with our experimental data. Essentially, the limit represents the small
strain aspect of Hooke's law upon which JKR theory is based. 
Thus, for material design considerations where strain is not known as a \emph{priori}, this limit in terms of elastoadhesive parameter can act as a crucial guideline.
Furthermore, using theoretical modeling for contact of rigid spheres on soft substrates, Zhang et al. \cite{zhang2019refined} made similar observations whereby JKR theory was found to be valid for $ E^{*}R_{0}/w \geq 100$ and exhibited significant deviations for $ E^{*}R_{0}/w \leq 10$. However, for their chosen contacting pairs, effects of solid surface tension was hypothesized to be dominant below the above mentioned limit whereas for the present study, we observe onset of wetting-like interactions. 

To physically interpret our experimental observation beyond just scaling laws, we revisit conventional contact theories relying on minimization of total energy (elastic, adhesion, capillary) of the contact system with respect to either contact radius $a$ or indentation depth $\delta$. For the small-to-large deformations present in the current work, the appropriate elastic
energy is the solution of the Boussinesq problem \cite{maugis1995extension,salez2013adhesion}, $U_{\rm el} \sim   E^{*} {R_{0}}^{3}  \int_{0}^{a/R_{0}} \left[1/2 - {(1+x'^2)/4x'}\ln\left((1+x')/(1-x')\right)\right]^{2} \,dx' $, where the integral evaluates the strain $dx'$. Upon solving the integral using the approximation, $\ln[(1+x')/1-x')]\approx 2\tanh^{-1}x' \approx 2(x'+x'^3/3)$, we can express the elastic energy as, $U_{\rm el}\sim E^{*}{R_{0}}^{3}\left[{1/9}(a/R_{0})^{9} + {8/7}(a/R_{0})^{7}+{16/5}(a/R_{0})^{5}\right] $. Minimizing total energy, $U = U_{\rm el} + U_{\rm ad}$, where $U_{\rm ad} = -\pi w a^{2}$ is the adhesion energy, with respect to contact radius $a$, we obtain the following relation between strain and elasto-adhesive parameter,
\begin{equation}
\left(\frac{a}{R_{0}}\right)^{7}+8\left(\frac{a}{R_{0}}\right)^{5}+16\left(\frac{a}{R_{0}}\right)^{3} = \frac{72\pi w}{E^{*}R_{0}}.
\label{eqn:2}
\end{equation}
On numerically solving Eq.~\ref{eqn:2}, we observe that the proposed
model smoothly translates our experimental data from low to high strains (Fig.~\ref{fig:4}b), even beyond $a/R_{0} = 1$, diverging only at $E^{*} = 0$ (pure liquid case). Also, for Eq.~\ref{eqn:2}, only one
\emph{real solution} exists whereas the remaining six solutions are \emph{complex}. 
Further, we observe that for low strains, i.e., $a/R_{0} < 0.4$,
the last term on L.H.S of Eq.~\ref{eqn:2} becomes dominant and the proposed model recovers the classical JKR solution (Fig.~\ref{fig:4}b). In the energy minimization approach, often a capillary term is added accounting for the energy of the hydrogel spherical cap profile \cite{salez2013adhesion}: $U_{\rm cap} \approx 2\pi\gamma(a^2+b^2/2)$,
where $b$ is the vertical height of the hydrogel. For adhesive contacts, $b>>a$, $U_{\rm cap} \approx \pi\gamma b^2 $ and thus the
term vanishes upon partial derivative w.r.t. $a$. For contacts in the transition zone,
$a \approx b$, $U_{\rm cap} \approx 3 \pi\gamma a^{2}$ and thus 
$(\partial{U_{\rm cap}}/\partial{a}) \sim \pi\gamma a$. And for wettting like configuration, $a >> b$, $U_{\rm cap} \approx 2 \pi\gamma a^2 $ and thus $(\partial U_{\rm cap}/\partial a) \sim \pi\gamma a$. Hence, the contribution of $U_{\rm cap}$ is insignificant for $E^{*}R_{0}/w > 200$ 
and causes a slight downshift of the solution of Eq.~\ref{eqn:2} for  
$E^{*}R_{0}/w < 200$. 

Here, we note that as $a/R_{0} \rightarrow 0$, the present contact problem extends well beyond Hookean elasticity and may include additional factors like non-linear elasticity \cite{tatara1991compression} or plasticity \cite{maugis1984surface}. 
Additionally, surface tension effects (hydrogel, PDMS) are expected to contribute for $ E^{*}R_{0}/w \leq 10$. 
As mentioned previously, we observe onset of wetting like interactions for the softest hydrogel ($E_{1} = 0.0057\,{\rm kPa}$), where the contact radius is dependent on substrate wettability typical of liquids. 
As shown in Fig.~\ref{fig:4}, this behavior corresponds to $E^{*}R_{0}/w = 0.06$ suggesting that onset of wetting occurs when adhesion significantly dominates over elasticity. 
We note here that for simplicity, we have used $E^{*}R_{0}/w \approx 0.1$ as the bound for onset of wetting although $E^{*}R_{0}/w \approx 1$ is equally likely.
Eventually, the contribution from elasticity disappears in the liquid limit ($E^{*} = 0$), where adhesion induced contact is solely resisted by liquid (hydrogel) surface tension $\gamma$. Here, we can express the contact as $a/R_{0} \sim f(\gamma/w)$. 
For the simplest case, $a/R_{0} \approx \sqrt{w/2\gamma} \approx 1$ \cite{hui2015indentation}, which obeys our experimentally observed contact radius values for very soft/liquid hydrogels as well as water droplets on pristine PDMS substrates. In this liquid limit, elasticity effects will influence the wetting morphology only when the underlying substrate is sufficiently soft. The manifestation of this effect provides the \emph{wetting ridge} structure, localized at the three-phase contact line \cite{style2013universal,marchand2012contact,
jerison2011deformation}. This particular aspect is further discussed in the following section.

\subsection{Deformation foot and wetting ridge}
Interestingly, the contact radius is not the sole variable aiding our understanding of this transition problem. 
As mentioned previously, the hydrogels undergo a deformation close to the contact and exhibit a foot-like region (Fig.~\ref{fig:5}a). The foot diminishes as the hydrogel elasticity increases (or decreases) and resembles the spherical cap configuration of water droplets for the \emph{liquid hydrogel}. Here, we probe the evolution of the hydrogel foot-height to better understand the adhesion-wetting transition. 
Briefly, we recall that for rigid glass spheres ($E_{1} \approx \infty$) and liquid droplets ($E_{1} \approx 0$) on soft substrates, deformation creates a ridge like structure on the soft substrates below the contact line  \cite{butt2005force,style2013surface,kim2022capillary,
style2013universal,marchand2012contact,
jerison2011deformation,style2017elastocapillarity,park2014visualization} (Fig.~\ref{fig:5}a). However, for a finite elasticity $E_{1}$, the deformation forms a foot like structure into the hydrogel close to the contact line.
As shown in Fig.~\ref{fig:5}b-c, the foot height increases with decreasing $E^{*}R_{0}/w$ and eventually experiences an inflection point at $E^{*}R_{0}/w \approx 0.2 - 0.5$ (or equivalently, $E_{1} \approx 0.005\,{\rm kPa} - 0.1\,{\rm kPa}$). 
Below that threshold value, the deformation occurs in the soft substrates at the contact line. The commonly called \emph{wetting ridge} is observed for the liquid-like hydrogel ($\mu_{1} = 0.163\,{\rm Pa.s}$) on the soft substrates. As convention, we represent deformation in the hydrogel as +ve and that in the soft substrate as -ve in Fig.~\ref{fig:5}b-c. 
Further, we observe that the hydrogel with least elasticity, i.e., $E_{1} = 0.0057\,{\rm kPa}$ exhibits no \emph{foot} for some of the softer PDMS substrates whereas the liquid hydrogel exhibits no \emph{wetting ridge} for some of the stiffer PDMS substrates indicating a delicate dependence on $E^{*}$ at the inflection point. Lastly, in stark contrast to deformation in soft substrates, which obey $h \sim \gamma/E_{2}$ \cite{style2013universal,marchand2012contact,
jerison2011deformation}, here we observe that the hydrogel foot height can be represented using a scaling law, $h/R_{0} \sim \left(w/E^{*}R_{0}\right)^{1/2}$ (Fig.~\ref{fig:5}b): a dependence with $E^{-1/2}$ rather than $E^{-1}$. 
However, a more rigorous theoretical analysis to interpret the
parabolic nature of foot height $h$ across the entire parameter space is beyond the scope of the present work.

\section{Conclusions}
Through careful experimentation of hydrogels, water droplets and rigid glass spheres on soft PDMS and rigid glass substrates, 
we have highlighted the transition from contacting surfaces to wetting of liquids. Our observations indicate that the transition can be represented as an extensive zone using the characteristic elasto-adhesive parameter $E^{*}R_{0}/w$ which accounts for elasticity of both the contacting pairs as well the work of adhesion between the two. From the analysis of contact/wetting radius, we identify that the transition region has an upper bound, i.e., $E^{*}R_{0}/w \approx 200$, beyond which JKR theory can adequately explain contacting surfaces. As a lower bound,
$E^{*}R_{0}/w \approx 0.1$ is indicative of a transition to wetting of liquids. At the same time, we highlight that the crossover of the contact/wetting radius from adhesion to wetting can be expressed using a higher order polynomial in strain $a/R_{0}$ with respect to $E^{*}R_{0}/w$, which recovers the JKR theory for small strains. Comparison with existing literature data  
yields satisfactory agreement.
Lastly, analysis of the foot-like deformation region of the hydrogel provides us with a crucial insight: as $E^{*}R_{0}/w$ approaches zero, the deformation in the hydrogel (top contacting pair) tends to zero and instead onset of deformation in the soft substrate (bottom contacting pair), i.e., the wetting ridge occurs.


\section{Experimental Section}
\subsection{Fabrication of hydrogels and soft substrates}
The chemicals used in this work were purchased from Sigma-Aldrich, if not specified. To synthesize hydrogels, we used acrylamide (AAm) as a monomer, N,N’-Methylene-bis-acrylamide (BIS) as the crosslinker, and 2,4,6-tri-methyl benzoyl-diphenylphosphine oxide (TPO) nanoparticle as the initiator. TPO nanoparticles were synthesized by dissolving 2.5 wt.\,\% of diphenyl (2,4,6-trimethylbenzoyl) phosphine oxide 
($\rm {M_{w}} = 348.48$), 3.75 wt.\,\% of polyvinylpyrrolidone, and 3.75 wt.\,\% of dodecyl surface sodium salt (SDS) in  DI water. After mixing the ingredients with a sonicator for 5 min at 95$^{\circ}$C, we obtained 10 wt.\,\% of TPO nanoparticles in DI water. Pregel solutions were prepared by diluting the monomer (2.5 – 13 wt.\%), 1 wt.\% of the crosslinker (based on the monomer), and 2.5 wt. \% of TPO nanoparticles (based on the monomer) in 0.5\,mM NaIO$_{\rm {4}}$ solution (oxygen scavenger). The hydrogel beads were prepared by suspending 4\,$\mu$L of the pregel solution in a beaker containing n-octane and a silicone oil (phenylmethylsiloxane-dimethylsiloxane copolymer, 500\,cSt, Gelest). The volume ratio between n-octane and the silicone oil is maintained as 1:2. 
Spherical shapes of the hydrogel were achieved due to the density gradient between n-octane (density $\rho$\,=\,0.71 g/$\rm {cm^{3}}$) and the silicone oil ($\rho$\,=\,1.08 g/$\rm {cm^{3}}$). 
Each pregel solution was exposed to UV light ($\sim$\,365 nm) for 20 min. 
The cured hydrogel beads were washed with heptane multiple times before each use.

Soft substrates were prepared using polydimethylsiloxane (PDMS, Sylgard 184 and Sylgard 527). 
First, the base and curing agent of Sylgard 184 and 527 were mixed in the weight ratio of 10:1 and 1:1, respectively as per manufacturer specifications. 
Consequently, each prepolymer of Sylgard 184 (10:1) and Sylgard 527 (1:1) were mixed together in different weight ratios of 1:0, 1.5:1, 1:1, 1:4, 1:15, and 0:1 to modulate the elastic properties. 
The mixture was then thoroughly stirred, vacuumed to eliminate trapped bubbles and cured at 85$^\circ$C for 12\,hours and then kept at room temperature for 2\,hours before each use. 
Using this technique, we prepared 2 mm-thick soft substrates by controlling the prepolymer volume in a petri dish.  
Mixing Sylgard 184 (10:1) and 527 (1:1) maintained the stoichiometry of each PDMS while decreasing the elastic modulus from a few MPa to kPa. 
This mixture is often used for studying biological tissue mechanics due to the fine control of mechanical strength  \cite{palchesko2012development}. 

\subsection{Rheology of hydrogels and soft substrates}
For rheology measurements, each PAAm hydrogel and silicone gel was polymerized in a 60 mm-diameter-petri dish 
with a thickness of 2 mm. 
Detailed fabrication steps are illustrated in the previous section. The cured hydrogels and silicone gels were cut into 25 mm-diameter using a cutter. The shear storage and loss modulus of the materials were measured by performing a frequency sweep test on a dynamic shear rheometer (AR 2000, TA Instruments) from 0.01 to 100 Hz at a strain rate of 1\,\% and a normal force of 1\,N. 
A constant temperature was maintained at 25$^{\circ}$C, 
and the test adopter is a 25 mm diameter plate. 
The measurement is taken after waiting for 10 min to stabilize the polymer. Each measurement was repeated three times. Further, rheology measurements were also performed for plasma-treated PDMS substrates and no noticeable change in elasticity was observed with their bare (no plasma) counterpart.

The shear viscosity of PAAm 2.5 wt.\% was analyzed from 0.01 to 100 Hz using a dynamic shear rheometer (Kinexus Rotational Rheometer, Malvern Instruments) at 25${^\circ}$C. Since the monomer ratio was very low, the cured PAAm 2.5 wt. \% was still liquid, having less than 1\,Pa of the shear storage modulus. As a result, we measured the shear viscosity of PAAm 2.5 wt.\% instead of the shear elasticities like other hydrogels. A cup and bob geometry (C14:CP14) was used, with a rotating cylinder inserted inside the cup. The cured hydrogel was poured into the geometry, and then the temperature was equilibrated for 5\,min before testing. The shear rates from 0.01 to 100\,Hz were applied. The measurement was repeated three times. 

\subsection{Surface tension of hydrogels}
The surface tension of each prepolymer solution was measured using the pendant drop method under ambient condition of 25$^{\circ}$C on a drop-shape analyzer (Kruss, DSA30). 
The droplet volume was increased using at the tip of a needle (diameter: 0.8 mm) right before it dropped. 
The surface tension was measured based on the Young-Laplace equation programmed on the ADVANCE software (Kruss, DSA30). 
Each measurement was repeated three times. 
The surface energy of silicone gel was assumed to be equal to the measured surface tension of silicone prepolymer because of the small strain.   

\subsection{AFM Measurements}
Further, we examined the effect of plasma treatment on PDMS surface roughness using atomic force microscopy (AFM) (MFP-3D BIO, Asylum Research). Each PDMS substrate was prepared as mentioned earlier, and treated with air plasma for 5 min. For comparison, the surface roughness of pristine PDMS and bare glass slide were also measured. The average root mean square height ($R_{\rm q}$) and arithmetic mean roughness ($R_{\rm a}$) of a bare glass slide are 0.54\,$\pm$\,0.01\,nm and 0.43\,$\pm$\,0.01\,nm, respectively. 
On average, $R_{\rm q}$ and $R_{\rm a}$ of PDMS are 0.48\,$\pm$\,0.05\,nm and 0.31\,$\pm$\,0.03\,nm before air-plasma, respectively. After air plasma treatment, $R_{\rm q}$ and $R_{\rm a}$ values increase marginally as 0.66\,$\pm$\,0.03\,nm and 0.52\,$\pm$\,0.03\,nm, respectively.

\subsection{FTIR Measurements}
Any potential diffusion effect of PAAm hydrogel into PDMS substrate was examined under a Tensor 27 Attenuated Total Reflectance-Fourier Transformed Infrared (ATR-FTIR) spectrometer (Bruker). The measurements were obtained at a 1.92 $\rm {cm^{-1}}$ resolution from 124 scans. We placed 2\,mm thick, cured hydrogel films ($E_{1} = 106.65\,{\rm kPa}$) on 2\,mm thick PDMS substrates ($E_{2} = 3\,{\rm kPa}$ and $E_{2} = 6855\,{\rm kPa}$). 
After 1 hour, we removed the hydrogel films and cut the PDMS substrates to scan the cross-section of the PDMS. 
The FTIR spectra were collected at the nearest section of the surface where the hydrogel film was in contact. 
Regardless of the hydrogel contact, all the spectra overlap (Fig.~1d of the main text). 
As shown in Fig.~1d of the main text, the main absorption peaks of PDMS were observed at wavenumber of 
2964\,$\rm {cm^{-1}}$ (C-H stretching in CH$_{\rm {3}}$), 1257\,$\rm {cm^{-1}}$ (CH$_{\rm {3}}$ symmetric bending in Si-CH$_{\rm {3}}$), 1251 $\rm {cm^{-1}}$ and 1010 $\rm {cm^{-1}}$ (Si-O-Si), and 787 $\rm {cm^{-1}}$ (CH$_{\rm {3}}$ rocking in Si-CH$_{\rm {3}}$). 
The typical absorption peaks of PAAm hydrogel were not observed: at wavenumber of 1625 $\rm {cm^{-1}}$ (C-O stretching), 1599 $\rm {cm^{-1}}$ (N-H bending), and 1452 $\rm {cm^{-1}}$ and 1325 $\rm {cm^{-1}}$ (the scissoring and twisting vibrations of CH$_{\rm {2}}$). 
Also, the spectra does not manifest the IR absorption of water at wavenumber of 3500 $\rm {cm^{-1}}$ (O-H stretch) and 1635 $\rm {cm^{-1}}$ (O-H-O scissors bending). 

\subsection{Imaging experiments}
The side view of the static contact was imaged using FASTCAM Mini AX high-speed camera (Photron) coupled to the Resolv4K lens (Navitar). A 4x objective lens (Olympus) was additionally attached to attain the spatial resolution of 3 – 4\,$\mu$m/pixel. For cases, where higher magnification was required, a 10x long-working distance objective lens (Optem Inc.) was used providing a spatial resolution of 2\,$\mu$m/pixel.
Hydrogels were placed on the silicone gels using a PTFE-coated stainless steel dispensing needle. 
The captured side view images were subsequently processed using custom MATLAB routines and imageJ. 
The contact line was located where the gray scale intensity presented the maximum slope. Within the detected boundary of the hydrogel profile, the maximum circle was fitted, and the corresponding radius, wetting foot, wetting ridge height, macroscopic contact angle, and contact radius were measured. 
For each pair of hydrogels and soft substrates, three to five static measurements were performed. Since the prepared hydrogels have high water content, varying between 89.6$\%$ for hydrogel with 13.0\,wt.$\%$ monomer (AAm) to 97.6$\%$ for hydrogel with 2.5\,wt.$\%$ monomer (AAm), care was taken to perform imaging as soon as equilibrium configuration was reached to minimize dehydration of hydrogels. 
Here, we note that even though the water fraction is high in the prepared hydrogels, the water inside the hydrogel offers negligible force transmission and thus presents negligible contribution to adhesion \cite{yang2020hydrogel}. Consequently, although the crosslinked polymers are minority constituents in the hydrogel, they contribute to the adhesion with the other material present. 

As control experiments,
static measurements of water droplets and glass beads were also performed on bare glass slides and the soft substrates. 
For static configuration where the side view shadowgraphy provides inadequate resolution or clarity (for e.g., glass sphere contacts), we use bottom-view bright-field measurements at a magnification of 10x providing a resolution of 1\,$\mu$m/pixel. This is due to the fact that close to 180 contact angles exhibited by rigid glass spheres on PDMS as well as on glass slides pose significant restrictions to side-view imaging since the contact region appears to merge with its own shadow. This feature is often encountered in early time droplet spreading studies and provides an overestimation of extracted contact radius. Thus, bottom view imaging is employed to accurately pinpoint the three-phase contact line and extract the correct contact radius \cite{eddi2013short,mitra2016understanding}. 

\subsection{Adhesion measurements}
For measuring the work of adhesion between hydrogel and PDMS,
we used a cantilever based force probe \cite{shyam2023universal}. 
A polymeric capillary tube of diameter 410\,$\mu$m and spring constant, $k = 305\pm6.1\,{\rm nN/\mu m}$ is used as the cantilever probe. A hydrogel sphere/droplet was attached to the tip of the probe and the PDMS substrate (affixed to a linear actuator) was made to approach the probe at a prescribed velocity of 0.1 mm/s. Once contact was established, there was a hold time of around 15 s after which the PDMS substrate was made to retract. The adhesion induced interaction between PDMS and the hydrogel probe causes deflection $x$ of the cantilever. 
Consequently, the maximum deflection $\Delta x$ is measured and the corresponding peak adhesion force is calculated using, $F = k\Delta x$ . The work of adhesion is calculated using the relation for critical pull-off force. For purely JKR-type contacts, the critical pull off force can be expressed as, $F_{\rm c,JKR} = 3\pi R_{0} w/2$ \cite{johnson1971surface,prokopovich2011adhesion}. However, for the present case of different hydrogel contacts, we have added contributions to the above expression. The first added contribution comes from the capillary force from the hydrogel foot \cite{butt2005force}, $F_{\rm f} \approx 2\pi R_{0}\gamma\left(\cos{\theta^{*}}_{\rm L}+\cos{\theta^{*}}_{\rm R}\right) \approx 4\pi R_{0}\gamma \cos\theta^{*} $, where ${\theta^{*}}_{\rm L}$ and ${\theta^{*}}_{\rm R}$ are the foot contact angles. 
The second contribution comes from the capillary force from the spherical cap profile \cite{salez2013adhesion}, $F_{\rm c} \approx \pi \gamma (a^2 + b^2)/b $, where $b$ is the vertical height of the hydrogel.
Thus, the critical force becomes, $F \approx 4\pi R_{0}\gamma \cos\theta^{*} + \pi \gamma (a^2 + b^2)/b + 3\pi R_{0} w/2$. 
For example, for the experiment with PAAm 13\% on soft PDMS, we obtain the peak (pull-off) force $F = 1.6\,{\rm mN}$.
Consequently, using $\gamma = 61.9\,{\rm mN/m}$, $a \approx 0.48\,{\rm mm}$, $b \approx 1.97\,{\rm mm}$, $R_{0} \approx 1\,{\rm mm}$, and $\theta^{*} \approx 70^{\circ}$,
we calculate $w \approx 128\,{\rm mN/m}$. Incidentally, the calculated adhesion force is close to that obtained using $w \approx 2\gamma$, an assumption extensively used in existing literature \cite{butt2005force,liu2006deformation, kim2022capillary}.

\medskip
\textbf{Supporting Information} \par 
Supporting Information is available from the Wiley Online Library or from the author.

\medskip
\textbf{Acknowledgements} \par 
The authors are grateful to Lukas Bauman (Department of Chemical Engineering, University of Waterloo) for assistance with hydrogel preparation and Prof. Michael K.C. Tam (Department of Chemical Engineering, University of Waterloo) for providing the dynamic shear rheometer. A-R.K. acknowledges financial support from Waterloo Institute for Nanotechnology, University of Waterloo, in the form of Nanofellowship 2021. B.Z. acknowledges the support of NSERC RGPIN-2019-04650 and RGPAS-2019-00115. S.M., S.S., and S.K.M acknowledges the support of Discovery Grant (NSERC, RGPIN-2019-004060).

\medskip
\textbf{Conflict of interest} \par
The authors declare no competing interests.
%
\bibliographystyle{MSP}
\bibliography{ref}

\providecommand{\noopsort}[1]{}\providecommand{\singleletter}[1]{#1}%
\begin{thebibliography}{10}
\providecommand{\url}[1]{\texttt{#1}}
\providecommand{\urlprefix}{URL }

\bibitem{matsuzaki2014high}
T.~Matsuzaki, G.~Sazaki, M.~Suganuma, T.~Watanabe, T.~Yamazaki, M.~Tanaka,
  S.~Nakabayashi, H.~Y. Yoshikawa,
\newblock \emph{J. Phys. Chem. Lett.} \textbf{2014}, \emph{5}, 1 253.

\bibitem{zhou20233d}
T.~Zhou, H.~Yuk, F.~Hu, J.~Wu, F.~Tian, H.~Roh, Z.~Shen, G.~Gu, J.~Xu, B.~Lu,
  et~al.,
\newblock \emph{Nat. Mater.} \textbf{2023}, 1--8.

\bibitem{alici2015towards}
G.~Alici,
\newblock \emph{Expert Rev. Med. Devices} \textbf{2015}, \emph{12}, 6 703.

\bibitem{yang2020hydrogel}
J.~Yang, R.~Bai, B.~Chen, Z.~Suo,
\newblock \emph{Adv. Func. Mater.} \textbf{2020}, \emph{30}, 2 1901693.

\bibitem{stricher2015met}
A.~Stricher, R.~Rinaldi, C.~Barr{\`e}s, F.~Ganachaud, L.~Chazeau,
\newblock \emph{RSC Adv.} \textbf{2015}, \emph{5}, 66 53713.

\bibitem{yuk2016skin}
H.~Yuk, T.~Zhang, G.~A. Parada, X.~Liu, X.~Zhao,
\newblock \emph{Nat. Commun.} \textbf{2016}, \emph{7}, 1 12028.

\bibitem{shintake2018soft}
J.~Shintake, V.~Cacucciolo, D.~Floreano, H.~Shea,
\newblock \emph{Adv. Mater.} \textbf{2018}, \emph{30}, 29 1707035.

\bibitem{zhang2021inchworm}
Y.~Zhang, D.~Yang, P.~Yan, P.~Zhou, J.~Zou, G.~Gu,
\newblock \emph{IEEE Trans. Rob.} \textbf{2021}, \emph{38}, 3 1806.

\bibitem{fujii2002pdms}
T.~Fujii,
\newblock \emph{Microelectron. Eng.} \textbf{2002}, \emph{61} 907.

\bibitem{lamberti2014pdms}
A.~Lamberti, S.~L. Marasso, M.~Cocuzza,
\newblock \emph{RSC Adv.} \textbf{2014}, \emph{4}, 106 61415.

\bibitem{kim2019soft}
J.~Kim, E.-F. Chou, J.~Le, S.~Wong, M.~Chu, M.~Khine,
\newblock \emph{Adv. Healthcare Mater.} \textbf{2019}, \emph{8}, 13 1900109.

\bibitem{guo2016highly}
J.~Guo, X.~Liu, N.~Jiang, A.~K. Yetisen, H.~Yuk, C.~Yang, A.~Khademhosseini,
  X.~Zhao, S.-H. Yun,
\newblock \emph{Adv. Mater.} \textbf{2016}, \emph{28}, 46 10244.

\bibitem{demirci2022elastomer}
G.~Demirci, M.~J. Nied{\'z}wied{\'z}, N.~Kantor-Malujdy, M.~El~Fray,
\newblock \emph{Polymers} \textbf{2022}, \emph{14}, 9 1822.

\bibitem{grosjean2022bioresorbable}
M.~Grosjean, S.~Ouedraogo, S.~D{\'e}jean, X.~Garric, V.~Luchnikov, A.~Ponche,
  N.~Mathieu, K.~Anselme, B.~Nottelet,
\newblock \emph{ACS Appl. Mater. Interfaces} \textbf{2022}, \emph{14}, 38
  43719.

\bibitem{le2017wearable}
P.~Le~Floch, X.~Yao, Q.~Liu, Z.~Wang, G.~Nian, Y.~Sun, L.~Jia, Z.~Suo,
\newblock \emph{ACS Appl. Mater. Interfaces} \textbf{2017}, \emph{9}, 30 25542.

\bibitem{hertz1882beruhrung}
H.~Hertz,
\newblock \emph{J. f{\"u}r die reine und Angew. Math.} \textbf{1882},
  \emph{92}, 156-171 22.

\bibitem{johnson1971surface}
K.~L. Johnson, K.~Kendall, A.~D. Roberts,
\newblock \emph{Proc. Royal Soc. Lon. A.} \textbf{1971}, \emph{324}, 1558 301.

\bibitem{derjaguin1975effect}
B.~V. Derjaguin, V.~M. Muller, Y.~P. Toporov,
\newblock \emph{J. Colloid Interface Sci.} \textbf{1975}, \emph{53}, 2 314.

\bibitem{butt2005force}
H.-J. Butt, B.~Cappella, M.~Kappl,
\newblock \emph{Surf. Sci. Rep.} \textbf{2005}, \emph{59}, 1-6 1.

\bibitem{style2013surface}
R.~W. Style, C.~Hyland, R.~Boltyanskiy, J.~S. Wettlaufer, E.~R. Dufresne,
\newblock \emph{Nat. Commun.} \textbf{2013}, \emph{4}, 1 1.

\bibitem{liu2006deformation}
K.-K. Liu,
\newblock \emph{J. Phys. D: Appl. Phys.} \textbf{2006}, \emph{39}, 11 R189.

\bibitem{prokopovich2011adhesion}
P.~Prokopovich, V.~Starov,
\newblock \emph{Adv. Colloid Interface Sci.} \textbf{2011}, \emph{168}, 1-2
  210.

\bibitem{erath2010characterization}
J.~Erath, S.~Schmidt, A.~Fery,
\newblock \emph{Soft Matter} \textbf{2010}, \emph{6}, 7 1432.

\bibitem{maugis1995extension}
D.~Maugis,
\newblock \emph{Langmuir} \textbf{1995}, \emph{11}, 2 679.

\bibitem{pham2017elasticity}
J.~T. Pham, F.~Schellenberger, M.~Kappl, H.-J. Butt,
\newblock \emph{Phys. Rev. Mater.} \textbf{2017}, \emph{1}, 1 015602.

\bibitem{maugis1992adhesion}
D.~Maugis,
\newblock \emph{J. Colloid Interface Sci.} \textbf{1992}, \emph{150}, 1 243.

\bibitem{kim2022capillary}
A.-R. Kim, S.~K. Mitra, B.~Zhao,
\newblock \emph{J. Colloid Interface Sci.} \textbf{2022}, \emph{628} 788.

\bibitem{young1805iii}
T.~Young,
\newblock \emph{Phil. Trans. Royal Soc. Lon.} \textbf{1805}, , 95 65.

\bibitem{de1985wetting}
P.-G. De~Gennes,
\newblock \emph{Rev. Mod. Phys.} \textbf{1985}, \emph{57}, 3 827.

\bibitem{de2004capillarity}
P.-G. De~Gennes, F.~Brochard-Wyart, D.~Qu{\'e}r{\'e}, et~al.,
\newblock \emph{Capillarity and wetting phenomena: drops, bubbles, pearls,
  waves}, volume 315,
\newblock Springer, \textbf{2004}.

\bibitem{salez2013adhesion}
T.~Salez, M.~Benzaquen, {\'E}.~Rapha{\"e}l,
\newblock \emph{Soft Matter} \textbf{2013}, \emph{9}, 45 10699.

\bibitem{long2016effects}
J.~Long, G.~Wang, X.-Q. Feng, S.~Yu,
\newblock \emph{Int. J. Solids Struc.} \textbf{2016}, \emph{84} 133.

\bibitem{zhang2019refined}
L.~Zhang, C.~Ru,
\newblock \emph{J. Appl. Mech.} \textbf{2019}, \emph{86}, 5.

\bibitem{xu2014effects}
X.~Xu, A.~Jagota, C.-Y. Hui,
\newblock \emph{Soft Matter} \textbf{2014}, \emph{10}, 26 4625.

\bibitem{hui2015indentation}
C.-Y. Hui, T.~Liu, T.~Salez, E.~Raphael, A.~Jagota,
\newblock \emph{Proc. Royal Soc. A} \textbf{2015}, \emph{471}, 2175 20140727.

\bibitem{cao2014adhesion}
Z.~Cao, M.~J. Stevens, A.~V. Dobrynin,
\newblock \emph{Macromolecules} \textbf{2014}, \emph{47}, 9 3203.

\bibitem{rimai1989adhesion}
D.~Rimai, L.~DeMejo, R.~Bowen,
\newblock \emph{J. Appl. Phys.} \textbf{1989}, \emph{66}, 8 3574.

\bibitem{rimai1994adhesion}
D.~Rimai, L.~DeMejo, W.~Vreeland, R.~Bowen,
\newblock \emph{Langmuir} \textbf{1994}, \emph{10}, 11 4361.

\bibitem{ally2010interaction}
J.~Ally, E.~Vittorias, A.~Amirfazli, M.~Kappl, E.~Bonaccurso, C.~E. McNamee,
  H.-J. Butt,
\newblock \emph{Langmuir} \textbf{2010}, \emph{26}, 14 11797.

\bibitem{jensen2017strain}
K.~E. Jensen, R.~W. Style, Q.~Xu, E.~R. Dufresne,
\newblock \emph{Phys. Rev. X} \textbf{2017}, \emph{7}, 4 041031.

\bibitem{jensen2015wetting}
K.~E. Jensen, R.~Sarfati, R.~W. Style, R.~Boltyanskiy, A.~Chakrabarti, M.~K.
  Chaudhury, E.~R. Dufresne,
\newblock \emph{Proc. Nat. Acad. Sci.} \textbf{2015}, \emph{112}, 47 14490.

\bibitem{liu2021capillary}
L.~Liu, K.-K. Liu,
\newblock \emph{Colloids Surf. A} \textbf{2021}, \emph{611} 125828.

\bibitem{sokoloff2016effects}
J.~Sokoloff,
\newblock \emph{Langmuir} \textbf{2016}, \emph{32}, 1 135.

\bibitem{chakrabarti2018elastowetting}
A.~Chakrabarti, A.~Porat, E.~Rapha{\"e}l, T.~Salez, M.~K. Chaudhury,
\newblock \emph{Langmuir} \textbf{2018}, \emph{34}, 13 3894.

\bibitem{schwarz2013physics}
U.~S. Schwarz, S.~A. Safran,
\newblock \emph{Rev. Mod. Phys.} \textbf{2013}, \emph{85}, 3 1327.

\bibitem{zhao2017electronic}
S.~Zhao, R.~Zhu,
\newblock \emph{Adv. Mater.} \textbf{2017}, \emph{29}, 15 1606151.

\bibitem{gong2018matching}
Z.~Gong, S.~E. Szczesny, S.~R. Caliari, E.~E. Charrier, O.~Chaudhuri, X.~Cao,
  Y.~Lin, R.~L. Mauck, P.~A. Janmey, J.~A. Burdick, et~al.,
\newblock \emph{Proc. Nat. Acad. Sci.} \textbf{2018}, \emph{115}, 12 E2686.

\bibitem{li2008afm}
Q.~Li, G.~Y. Lee, C.~N. Ong, C.~T. Lim,
\newblock \emph{Biochem. Biophys. Res. Commun.} \textbf{2008}, \emph{374}, 4
  609.

\bibitem{regehr2009biological}
K.~J. Regehr, M.~Domenech, J.~T. Koepsel, K.~C. Carver, S.~J. Ellison-Zelski,
  W.~L. Murphy, L.~A. Schuler, E.~T. Alarid, D.~J. Beebe,
\newblock \emph{Lab Chip} \textbf{2009}, \emph{9}, 15 2132.

\bibitem{kopecek2009hydrogels}
J.~Kopecek,
\newblock \emph{J. Polym. Sci., Part A: Polym. Chem.} \textbf{2009}, \emph{47},
  22 5929.

\bibitem{style2013universal}
R.~W. Style, R.~Boltyanskiy, Y.~Che, J.~Wettlaufer, L.~A. Wilen, E.~R.
  Dufresne,
\newblock \emph{Phys. Rev. Lett.} \textbf{2013}, \emph{110}, 6 066103.

\bibitem{marchand2012contact}
A.~Marchand, S.~Das, J.~H. Snoeijer, B.~Andreotti,
\newblock \emph{Phys. Rev. Lett.} \textbf{2012}, \emph{109}, 23 236101.

\bibitem{jerison2011deformation}
E.~R. Jerison, Y.~Xu, L.~A. Wilen, E.~R. Dufresne,
\newblock \emph{Phys. Rev. Lett.} \textbf{2011}, \emph{106}, 18 186103.

\bibitem{joanny2001gels}
J.~Joanny, A.~Johner, T.~A. Vilgis,
\newblock \emph{Eur. Phys. J. E} \textbf{2001}, \emph{6}, 3 201.

\bibitem{bouillant2022rapid}
A.~Bouillant, P.~J. Dekker, M.~A. Hack, J.~H. Snoeijer,
\newblock \emph{Phys. Rev. Fluids} \textbf{2022}, \emph{7}, 12 123604.

\bibitem{tabor1977surface}
D.~Tabor,
\newblock In \emph{Plenary and invited lectures}, 3--14. Elsevier,
  \textbf{1977}.

\bibitem{maugis2000contact}
D.~Maugis,
\newblock \emph{Contact, adhesion and rupture of elastic solids}, volume 130,
\newblock Springer Science \& Business Media, \textbf{2000}.

\bibitem{tatara1991compression}
Y.~Tatara,
\newblock \emph{J. Eng. Mater. Technol.} \textbf{1991}, \emph{113} 285.

\bibitem{maugis1984surface}
D.~Maugis, H.~Pollock,
\newblock \emph{Acta Metall.} \textbf{1984}, \emph{32}, 9 1323.

\bibitem{style2017elastocapillarity}
R.~W. Style, A.~Jagota, C.-Y. Hui, E.~R. Dufresne,
\newblock \emph{Annu. Rev. Cond. Mat. Phys.} \textbf{2017}, \emph{8} 99.

\bibitem{park2014visualization}
S.~J. Park, B.~M. Weon, J.~S. Lee, J.~Lee, J.~Kim, J.~H. Je,
\newblock \emph{Nat. Commun.} \textbf{2014}, \emph{5}, 1 4369.

\bibitem{palchesko2012development}
R.~N. Palchesko, L.~Zhang, Y.~Sun, A.~W. Feinberg,
\newblock \emph{PloS One} \textbf{2012}, \emph{7}, 12 e51499.

\bibitem{eddi2013short}
A.~Eddi, K.~G. Winkels, J.~H. Snoeijer,
\newblock \emph{Phys. Fluids} \textbf{2013}, \emph{25}, 1 013102.

\bibitem{mitra2016understanding}
S.~Mitra, S.~K. Mitra,
\newblock \emph{Langmuir} \textbf{2016}, \emph{32}, 35 8843.

\bibitem{shyam2023universal}
S.~Shyam, S.~Misra, S.~K. Mitra,
\newblock \emph{J. Colloid Interface Sci.} \textbf{2023}, \emph{630} 322.

\end{thebibliography}

\begin{figure}
  \includegraphics[width=1\textwidth]{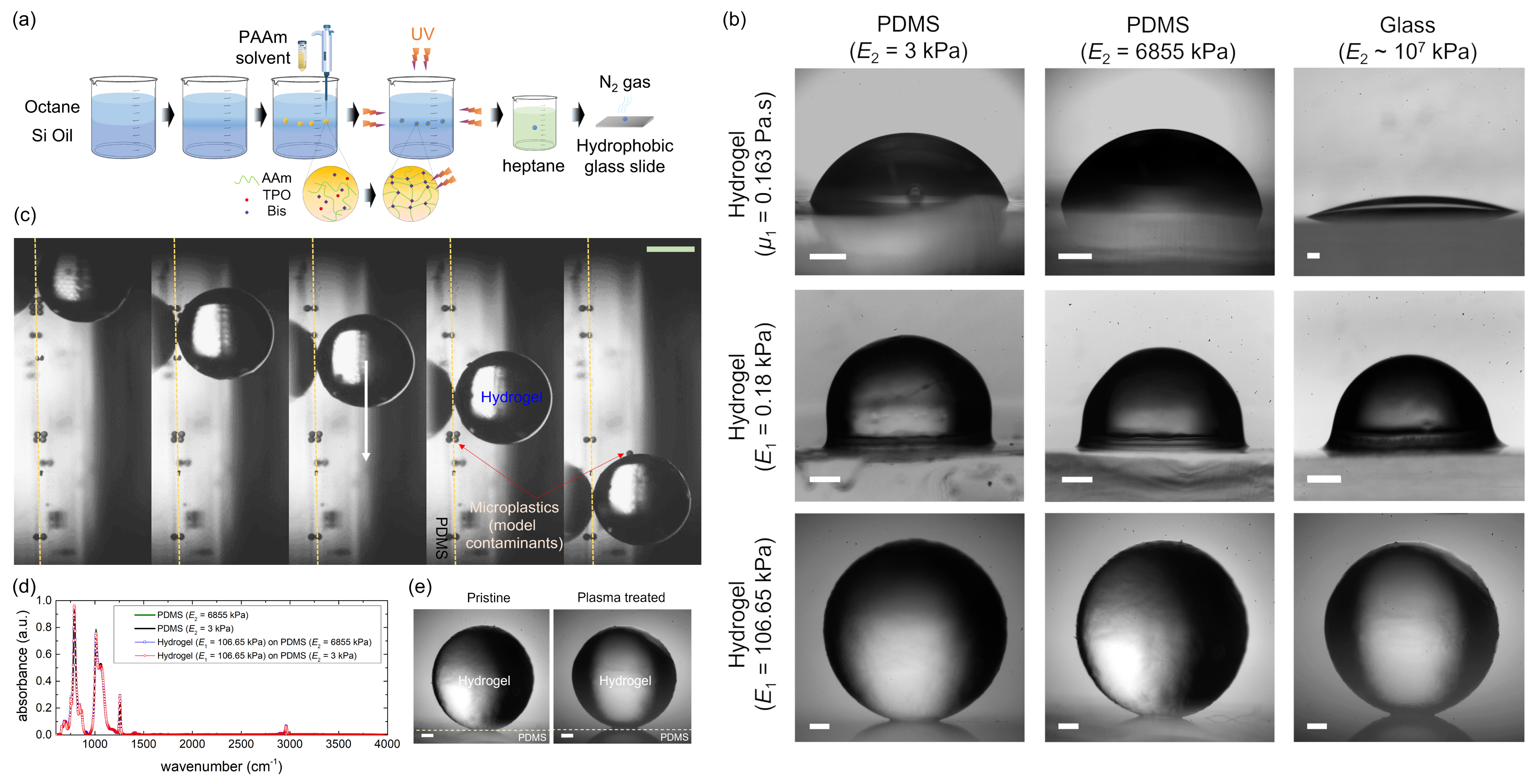}
  \caption{(a) Schematic showing the synthesis process of the Polyacrylamide (PAAm) hydrogels. Here, AAm: Acrylamide,
TPO: 2,4,6-tri-methyl benzoyl-diphenylphosphine oxide, and 
Bis: N,N’-Methylene-bis-acrylamide.  
(b) Equilibrium snapshots of hydrogels of varying elasticity ($E_{1}$) in contact with soft PDMS substrates ($E_{2} = 3\,{\rm kPa}$ and $E_{2} = 6855\,{\rm kPa}$)  and rigid glass substrates. The first row of images corresponds to the liquid-like hydrogels with $\mu = 0.163\,{\rm Pa.s}$ while the second and third row of images corresponds to hydrogels with Young's modulus $E_{1} = 0.18\,{\rm kPa}$ and $E_{1} = 106.65\,{\rm kPa}$, respectively. (c) Snapshots showing stiff hydrogels ($E_{1} = 392.80\,{\rm kPa}$) freely rolling down soft PDMS substrates ($E_{2} = 6855\,{\rm kPa}$) and in the process exhibiting cleaning of microplastics serving as model contaminants. The contaminants on the surface are not on the hydrogel path. The frames are separated by 1\,s. Scale bar represents 0.5\,mm. (d) FTIR spectra for pristine 2\,mm thick PDMS substrates ($E_{2} = 3\,{\rm kPa}$ and $E_{2} = 6855\,{\rm kPa}$) and the same substrates with hydrogel ($E_{1} = 106.65\,{\rm kPa}$) deposition. (e) Static snapshot of hydrogel ($E_{1} = 106.65\,{\rm kPa}$) on soft PDMS substrates ($E_{2} = 3\,{\rm kPa}$) with and without plasma treatment, still exhibiting non-wetting characteristics for the stiff hydrogel.   }
  \label{fig:1}
\end{figure}

\begin{figure}
  \includegraphics[width=1\textwidth]{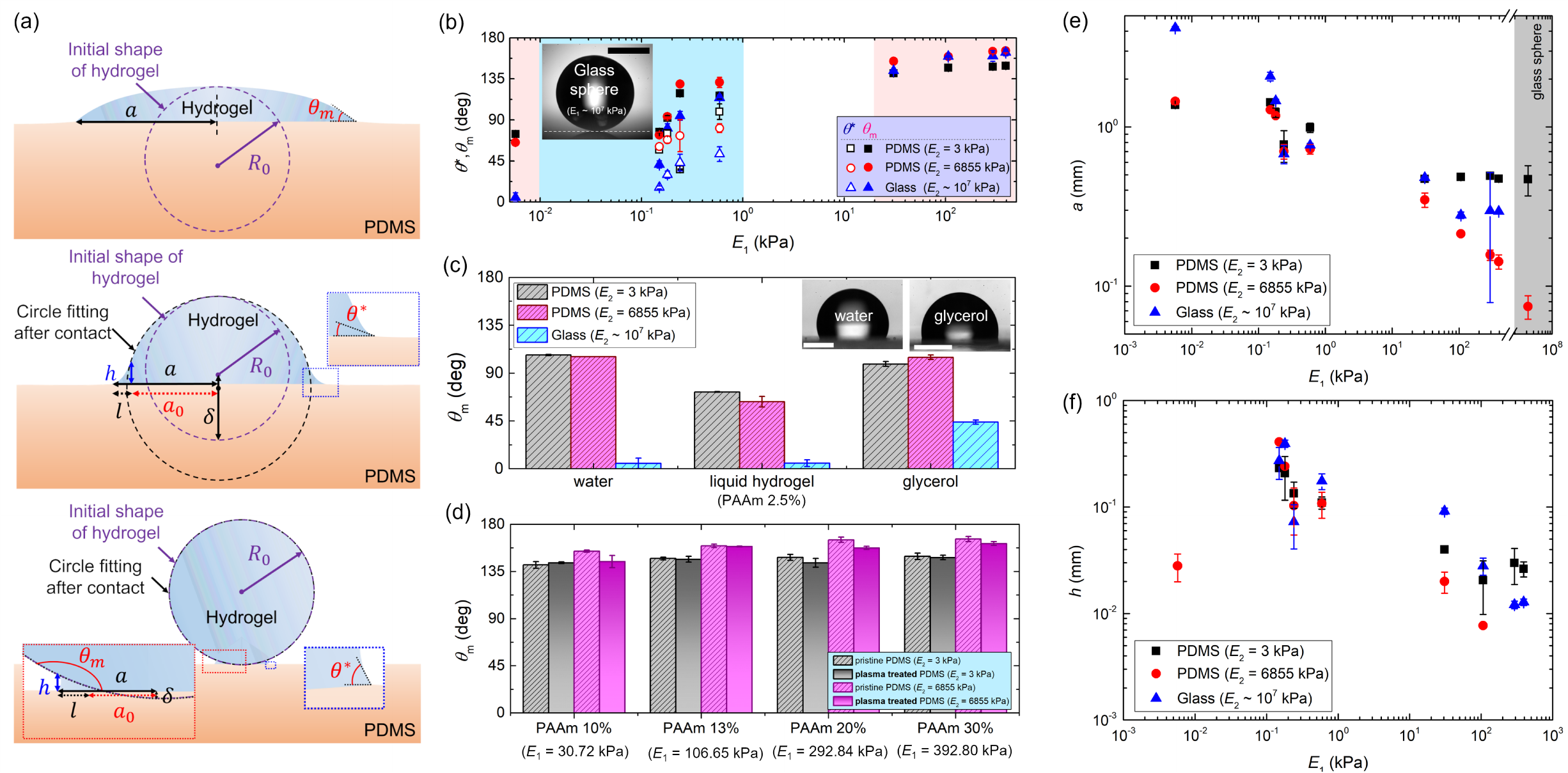}
  \caption{(a) Schematics of the different contact/wetting configuration of hydrogels on soft/rigid substrates. Here, $R_{0}$ and $a$ are the initial hydrogel radius and contact radius, respectively. $a_{0}$ and $\delta$ are the fitted {\emph{apparent}} contact radius and {\emph{apparent}} indentation depth, respectively. $h$ and $l$ are the foot-height and foot-length, respectively. $\theta^{*}$ is the contact angle of the foot while $\theta_{\rm m}$ is the macroscopic contact angle. (b) Variation of contact angle ($\theta^{*}$,$\theta_{\rm m}$) with hydrogel elasticity $E_{1}$ on pristine, soft PDMS ($E_{2} = 3\,{\rm kPa}$ and $E_{2} = 6855\,{\rm kPa}$) and rigid glass substrates. Inset shows the snapshot of a rigid sphere on a rigid glass substrate. Scale bar represents 0.5 mm. (c) Macroscopic contact angles $\theta_{\rm m}$ of \emph{liquid hydrogel} and common liquids like water and glycerol on pristine, soft PDMS ($E_{2} = 3\,{\rm kPa}$ and $E_{2} = 6855\,{\rm kPa}$) and rigid glass substrates. Insets show the snapshots of 1\,mm radius water and glycerol drops wetting rigid glass substrates. Scale bars represent 0.5 mm. (d) Variation of macroscopic contact angles $\theta_{\rm m}$ of relatively stiffer hydrogels on soft PDMS ($E_{2} = 3\,{\rm kPa}$ and $E_{2} = 6855\,{\rm kPa}$) substrates with (pristine) and without plasma treatment. (e)  Variation of contact radius $a$ with hydrogel elasticity $E_{1}$ on pristine, soft PDMS ($E_{2} = 3\,{\rm kPa}$ and $E_{2} = 6855\,{\rm kPa}$) and rigid glass substrates. The contact radius data for rigid glass spheres are also shown. (f) Variation of foot height $h$ with hydrogel elasticity $E_{1}$ on pristine, soft PDMS ($E_{2} = 3\,{\rm kPa}$ and $E_{2} = 6855\,{\rm kPa}$) and rigid glass substrates. }
  \label{fig:2}
\end{figure}

\begin{figure}
\centering
\includegraphics[width=1\textwidth]{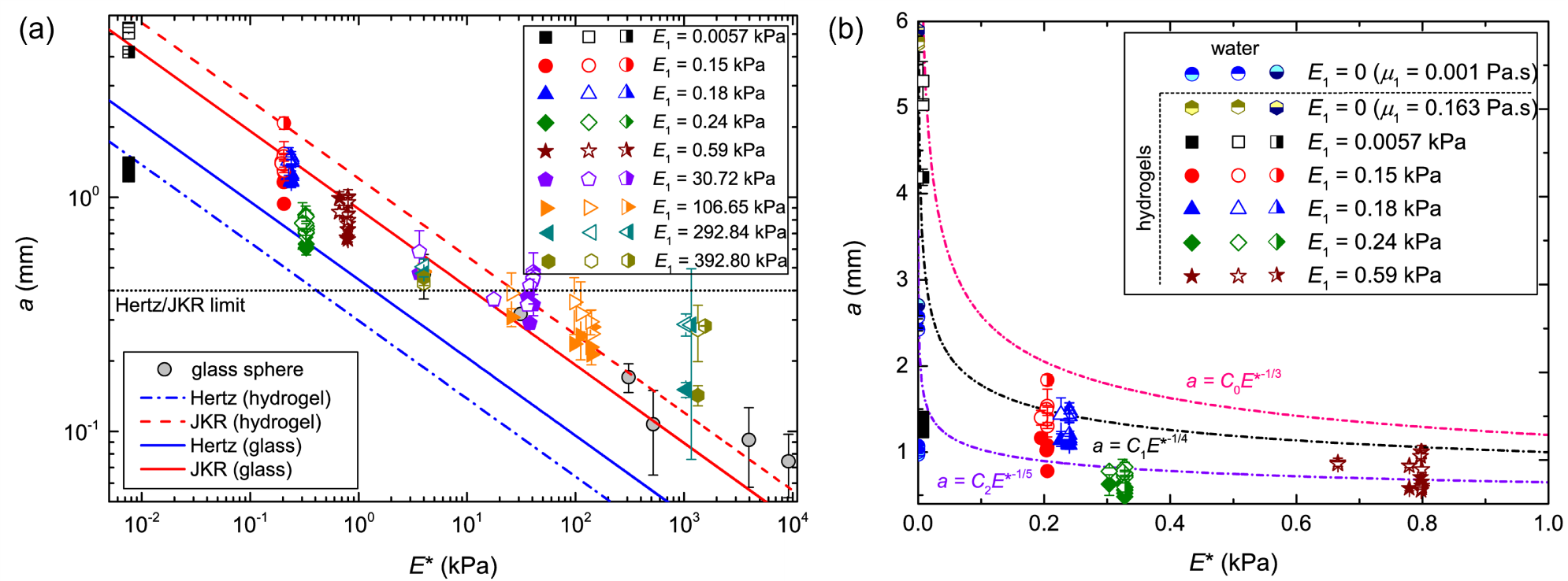}
\caption{(a)  Variation of contact radius $a$ with effective Young's modulus $E^{*}$ for hydrogels-on-PDMS, hydrogels-on-glass slide, and glass sphere-on-PDMS system. $E_{1}$ represents the elastic modulus of the hydrogels whereas PDMS elasticity $E_{2}$ is embedded in $E^{*}$. The closed symbols, open symbols and half-closed symbols represent hydrogels on pristine PDMS, hydrogels on plasma-treated PDMS and hydrogels on glass slides, respectively. The data for spherical glass spheres on PDMS (grey circles) is also shown. The Hertz and JKR theory (see main text) for both hydrogels (dashed lines) and glass spheres (solid lines) are plotted. The plots representing the theory are extended artificially beyond the Hertz/JKR limit of small strain, or equivalently $a \geq 0.4$ (see main text).
(b) Variation of contact radius $a$ with effective Young's modulus $E^{*}$ for hydrogels-on-PDMS as well as \emph{liquid hydrogel} and water droplets on PDMS (corresponds to $E^{*} = 0$). The dashed lines show the fitting $a \sim E^{*-1/3}$ (pink), $a \sim E^{*-1/4}$ (black), and $a \sim E^{*-1/5}$ (violet). The prefactors are $C_{0} = 1.2$, $C_{1} = 1.0$, and $C_{2} = 0.7$. The symbol coding is same as (a).
}
\label{fig:3}
\end{figure}

\begin{figure}
\centering
\includegraphics[width=1\textwidth]{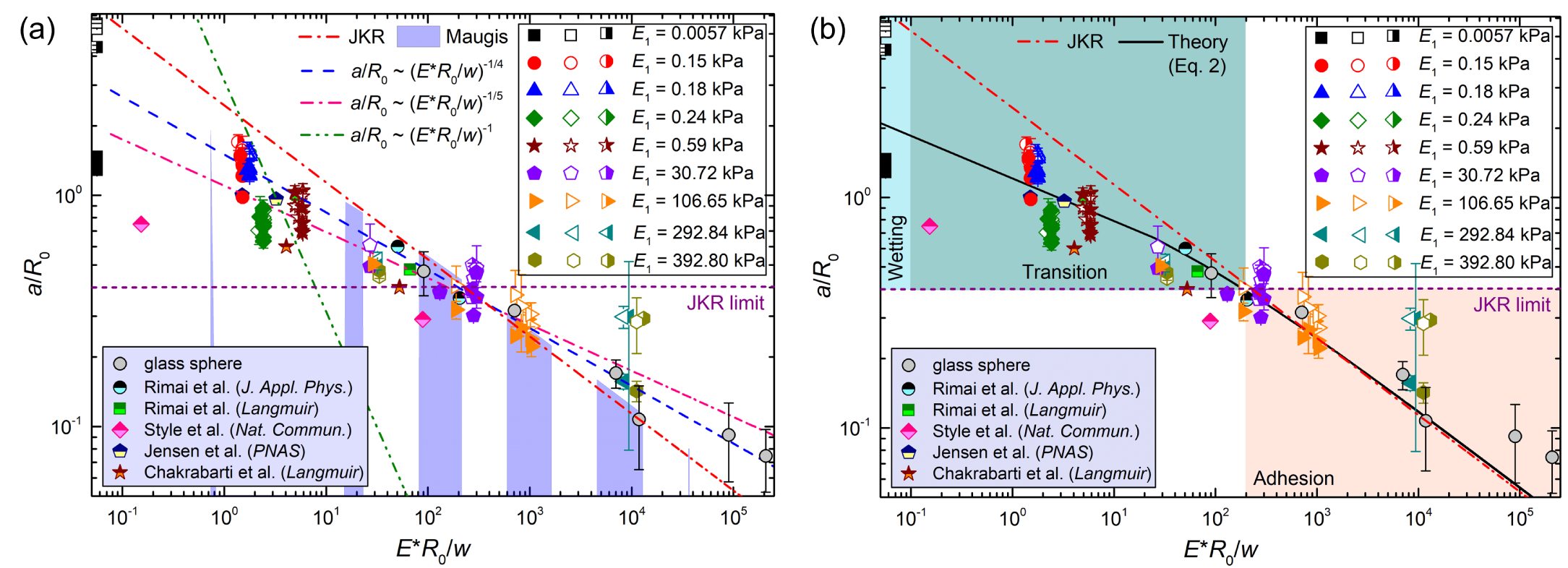}
\caption{(a) Variation of normalized contact/wetting radius $a/R_{0}$ with elasto-adhesive parameter $E^{*}R_{0}/w$ for hydrogels of different elasticity ($E_{1}$) on soft PDMS substrates of different elasticity($E_{2}$) as well as on rigid glass slides. The normalized contact radius data for rigid glass spheres on PDMS substrates are also shown. The contribution of $E_{2}$ is embedded in $E^{*}$. The closed symbols, open symbols and half-closed symbols represent hydrogels on pristine PDMS, hydrogels on plasma-treated PDMS and hydrogels on glass slides, respectively. The dashed red and dashed green lines represent the JKR and approximated Maugis model, respectively (see main text). The dashed blue and pink lines represent the scaling laws, $a/R_{0} \sim \left(E^{\ast}R_{0}/w\right)^{-1/4}$ and $a/R_{0} \sim \left(E^{\ast}R_{0}/w\right)^{-1/5}$, respectively. The blue shaded regions represent the standard Maugis model (Eq.~\ref{eqn:1}). 
Literature data \cite{style2013surface,rimai1989adhesion,rimai1994adhesion,
jensen2015wetting,chakrabarti2018elastowetting} are shown in the same plot. (b) Comparison of our experimental data (same as in (a)) with JKR theory (red dashed line) and the proposed model (Eq.~\ref{eqn:2}, black solid line). }
\label{fig:4}
\end{figure}

\begin{figure}
\centering
\includegraphics[width=1\textwidth]{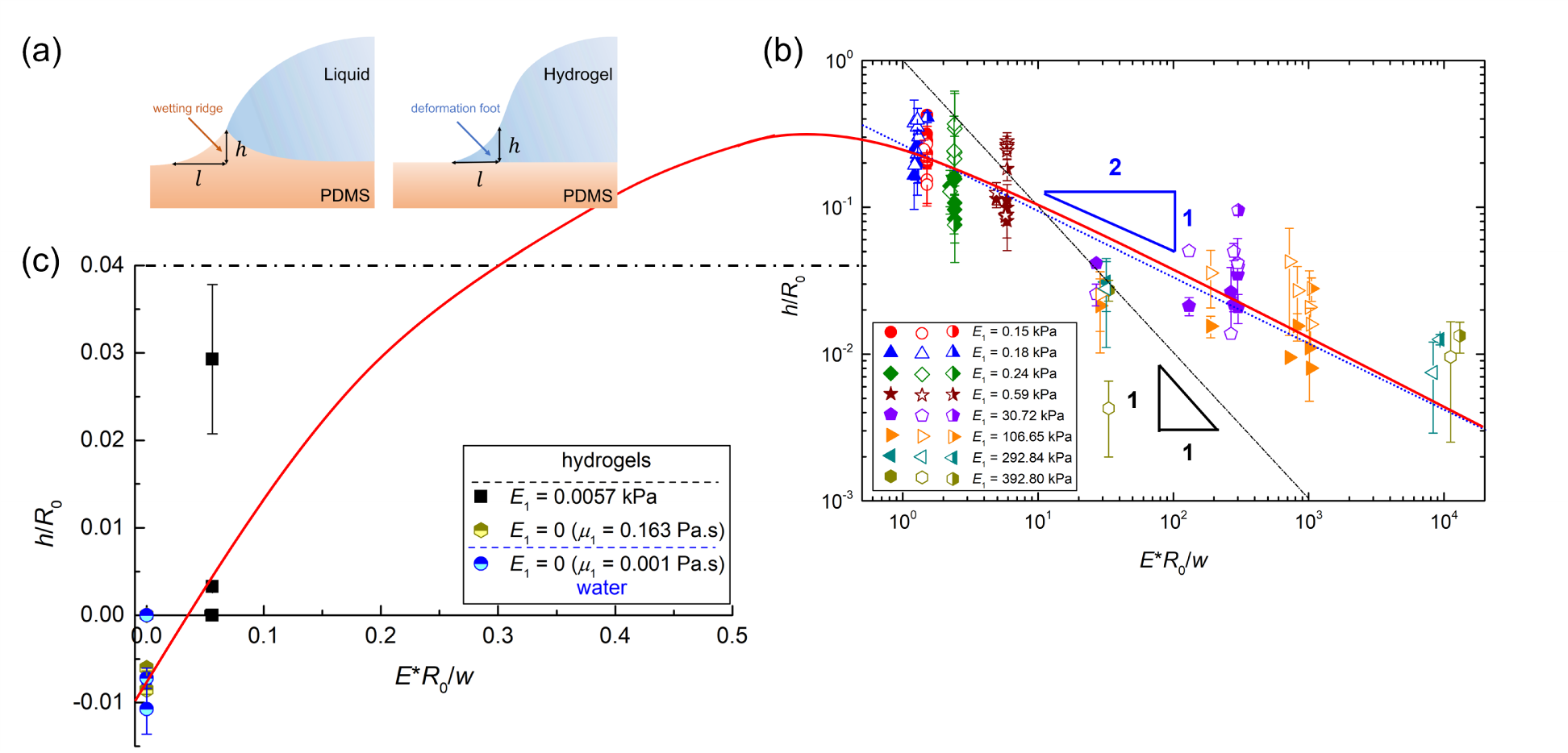}
\caption{(a) Schematics of liquid droplets on soft substrates forming a wetting ridge and hydrogels forming a distinct foot-like  deformation region. (b) Variation of normalized foot-height $h/R_{0}$ with  elasto-adhesive parameter $E^{*}R_{0}/w$ for hydrogels of different elasticity ($E_{1}$) on soft PDMS substrates of different elasticity($E_{2}$). The dashed black line represents the scaling law, $h/R_{0} \sim \left(E^{\ast}R_{0}/w\right)^{-1}$ whereas the dashed blue line represents the scaling law, $h/R_{0} \sim \left(E^{\ast}R_{0}/w\right)^{-1/2}$. The closed symbols, open symbols and half-closed symbols represent hydrogels on pristine PDMS, hydrogels on plasma-treated PDMS and hydrogels on glass slides, respectively. (c) Linear plot of variation of normalized foot-height $h/R_{0}$ with  elasto-adhesive parameter $E^{*}R_{0}/w$ for the hydrogel with the lowest elasticity, the \emph{liquid hydrogel} and water droplets. Deformation in hydrogel (foot) is along the +ve $h/R_{0}$ axis, while those in PDMS (wetting ridge) is along the -ve $h/R_{0}$ axis. The red spline is artificially drawn to highlight the transition from (b) to (c).  }
\label{fig:5}
\end{figure}



\end{document}